\documentclass[acmtog]{acmart}
\acmSubmissionID{528}

\usepackage{booktabs} 

\usepackage[ruled]{algorithm2e} 

\SetAlFnt{\small}
\SetAlCapFnt{\small}
\SetAlCapNameFnt{\small}
\SetAlCapHSkip{0pt}

\acmJournal{TOG}

\begin{document}
\title{STGA: Selective-Training Gaussian Head Avatars}

\author{Hanzhi Guo}
\email{hanzhiguo@bit.edu.cn}
\orcid{0009-0009-7496-5482}
\affiliation{%
  \institution{Beijing Institute of Technology}
  \city{Beijing}
  \country{China}
}

\author{Yixiao Chen}
\email{yxchengeorge@163.com}
\affiliation{%
  \institution{Beijing Institute of Technology}
  \city{Beijing}
  \country{China}
}

\author{Xiaonuo Dongye}
\email{dyxn@bit.edu.cn}
\affiliation{%
  \institution{Beijing Institute of Technology}
  \city{Beijing}
  \country{China}
}

\author{Zeyu Tian}
\email{tianty97@163.com}
\affiliation{%
  \institution{Beijing Institute of Technology}
  \city{Beijing}
  \country{China}
}

\author{Dongdong Weng}
\email{crgj@bit.edu.cn}
\affiliation{%
  \institution{Beijing Institute of Technology}
  \city{Beijing}
  \country{China}
}

\author{Le Luo}
\email{leluo1989@gmail.com}
\affiliation{%
  \institution{Pengcheng Laboratory}
  \city{Guangzhou}
  \country{China}
}

\renewcommand\shortauthors{Guo et al.}

\begin{abstract}
 We propose selective-training Gaussian head avatars (STGA) to enhance the details of dynamic head Gaussian. The dynamic head Gaussian model is trained based on the FLAME parameterized model. Each Gaussian splat is embedded within the FLAME mesh to achieve mesh-based animation of the Gaussian model. Before training, our selection strategy calculates the 3D Gaussian splat to be optimized in each frame. The parameters of these 3D Gaussian splats are optimized in the training of each frame, while those of the other splats are frozen. This means that the splats participating in the optimization process differ in each frame, to improve the realism of fine details. Compared with network-based methods, our method achieves better results with shorter training time. Compared with mesh-based methods, our method produces more realistic details within the same training time. Additionally, the ablation experiment confirms that our method effectively enhances the quality of details.
\end{abstract}
\vspace{-1.7cm}

%
%

\ccsdesc[500]{Computing methodologies~Point-based models}
\ccsdesc[500]{Computing methodologies~Rendering}
%
%

\vspace{-2cm}
\keywords{Gaussian splatting, facial animation, facial reenactment}

\begin{teaserfigure}
  \includegraphics[width=0.95\textwidth]{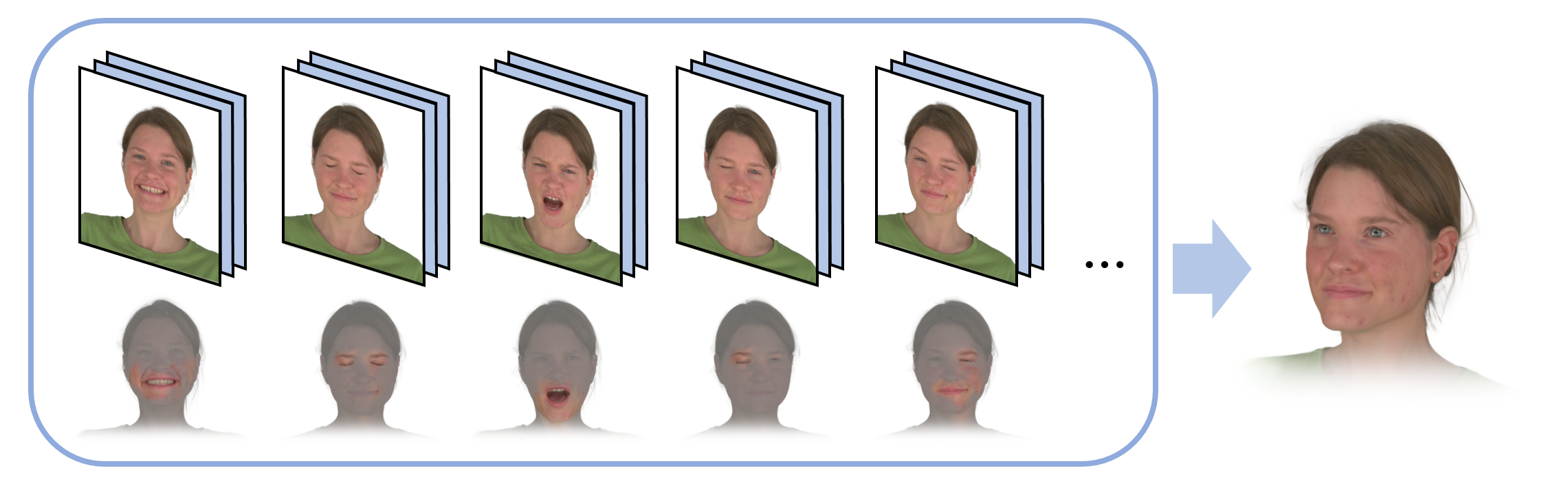}
  \vspace{-0.3cm}
  \caption{Our training method focuses on regions with significant facial expression changes during Gaussian optimization to enhance the details in the Gaussian heads and achieve more realistic driving results. The gray part represents the Gaussian splats that do not participate in training for each frame, while the colored part indicates the Gaussian splats involved in training for each frame.}
  \label{fig:teaser}
\end{teaserfigure}

\vspace{-0.6cm}
\maketitle

\vspace{-0.2cm}
\section{Introduction}

High-fidelity 3D avatar reconstruction is one of the important research directions in graphics and computer vision. The rapid development of technology provides unprecedented opportunities and challenges for the reconstruction of highly realistic drivable avatars. Reconstructing the avatars vividly and efficiently has always been a challenging problem.

According to the classification of 3D object representation methods, it can be divided into explicit representation methods and implicit representation methods. Commonly used explicit representation methods include mesh, point clouds, voxel, etc. Parametric mesh model~\cite{smpl, flame, mano, smplx}, is a typical explicit representation. Neural Radiance Fields (NeRF)~\cite{nerf}, as a representative method of implicit representation, combines deep learning and volume rendering. It can obtain highly realistic implicit scene models using multiview RGB images. Based on NeRF, many dynamic avatar representation methods~\cite{D_NeRF, nerfies, animatablenerf, humannerf, neuralactor} have been extended. These works achieve impressive results for dynamic avatar reconstruction. However, NeRF has several limitations. Its implicit representation makes the training results hard to edit and real-time rendering difficult to achieve. 

The recent 3D Gaussian Splatting~\cite{gaussian} method has demonstrated superior performance compared to NeRF. As a new explicit representation method, it has shown excellent results in novel view synthesis. 3D Gaussian Splatting offers advantages such as ease of training, explicit representation, and real-time rendering. Since the 3D Gaussian Splatting method was proposed, numerous research efforts have been conducted based on Gaussian representations~\cite{physgaussian, gaussian_grouping, GaussianReplacement}. And many new 3D avatar reconstruction methods based on it have emerged~\cite{gaussianavatars, splattingavatar, adaptive, bridging, splatface, gomavatar, humansplat}. The approach commonly used to drive Gaussian models is based on mesh models. By embedding Gaussian splats on the triangular surfaces of the mesh, the Gaussian model can be driven effectively. This method has proven effective for dynamic bodies and heads. However, creating a drivable Gaussian-based head requires capturing a large amount of training data to cover all expressions, ensuring that the trained model can perform realistic actions based on the driving data. However, there is a potential problem with this approach: if all Gaussian splats participate in each training round, it may lead to poor results in detail. This is because during training, the global optimization of Gaussian splats may sacrifice the precision of some details in exchange for the smoothness of the overall model and the minimum value of the global loss function.

To address this issue, we propose a novel training strategy. Employing the concept of embedding Gaussians into a mesh model~\cite{gaussianavatars}, our approach is driven by the FLAME~\cite{flame} model. The core of this method involves identifying regions within each frame that exhibit significant changes in facial expressions and subsequently optimizing them. First, each 3D Gaussian is embedded in a triangular surface of the mesh. Then, by comparing the differences between each frame's corresponding FLAME model and its neutral model, the extent of change in the triangular faces is quantified. Based on this quantification, triangular faces with significant changes are selected, and the Gaussian splats on these faces are chosen as the ones that require detailed optimization in that frame. During the training process, in addition to the local detail optimization iterations involving only a subset of the Gaussian splats, global optimization iterations involving all Gaussian splats are still retained. This alternating optimization approach ensures that the model maintains overall coherence while significantly enhancing the representation of regions with intricate details. Through this method, gaussian avatars can exhibit more realistic and natural details when performing complex actions, especially in cases involving subtle changes in facial expressions and muscle movements. 

The contributions of this work are as follows:
\vspace{-0.1cm}

\begin{itemize}
    \item We introduce a method for calculating the Gaussian splats to be optimized for each frame. By setting the pose parameters to zero, we identify the Gaussian splats requiring optimization in each frame based on the differences between the neutral mesh and the parameterized mesh.
    \item We propose a selective optimization method for dynamic Gaussian avatars,  where only the Gaussian splats in regions with stronger details are optimized in each frame. This approach enhances the details of Gaussian head models and improves the quality of the training results.
    \item We present both the reconstruction and reenactment results on the NeRSemble dataset, comparing our approach with mesh-based and network-based methods. The results of comparative experiments demonstrate that our method can get more realistic and detailed avatars.
\end{itemize}

\vspace{-0.2cm}
\section{Related Works}
The parametric model is the most commonly used explicit model of human 3D representation. Parametric models provide rich prior knowledge for human reconstruction, including the SMPL model~\cite{smpl} describing the human body, the MANO model~\cite{mano} describing the hand and the FLAME model~\cite{flame} describing the head. To obtain accurate model parameters from image data, some methods~\cite{emoca, mica} attempt to estimate the FLAME model of the target head from a single image, thereby acquiring the texture information of the target head~\cite{deca}. However, the topology of the parametric models limits the fineness of the model, and the quality of the texture image is difficult to achieve very fine. In addition, the parametric model also has certain limitations in describing details such as hair, beard and so on. These make the realism of explicit model-based reconstruction unable to reach a very high level.

In order to get rid of the limitations of the mesh model, researchers have explored the field of dynamic avatar reconstruction using a neural radiation field in recent years. In order to train the dynamic human neural radiation field, Animatable NeRF~\cite{animatablenerf} adopts the method of training the target human in the canonical space based on SMPL and rendering the driving result on the posed space. Neural actors~\cite{neuralactor} introduce texture information into neural radiation field training, but this method does not work well when expressing loose clothing and hair. Human NeRF~\cite{humannerf} introduces a non-rigid deformation network in attitude deformation, which can better simulate the flexible motion of clothing and other parts. Panohead~\cite{panohead} combines NeRF with GAN to generate realistic facial neural radiance field models from a single image. However, this method struggles to drive the resulting head effectively. Additionally, neural radiation field training requires a lot of computing power and time, and volume rendering algorithm limits the real-time output results. There are several ways to speed up NeRF training and rendering. These methods include replacing the previous model of the entire neural radiation field by using a large number of small MLPs~\cite{kilonerf}, using hash coding, or expressing the neural radiation field in octree form instead of network.~\cite{instant, plenoctrees}.

\begin{figure*}[h]
    \centering
    \includegraphics[width=\linewidth]{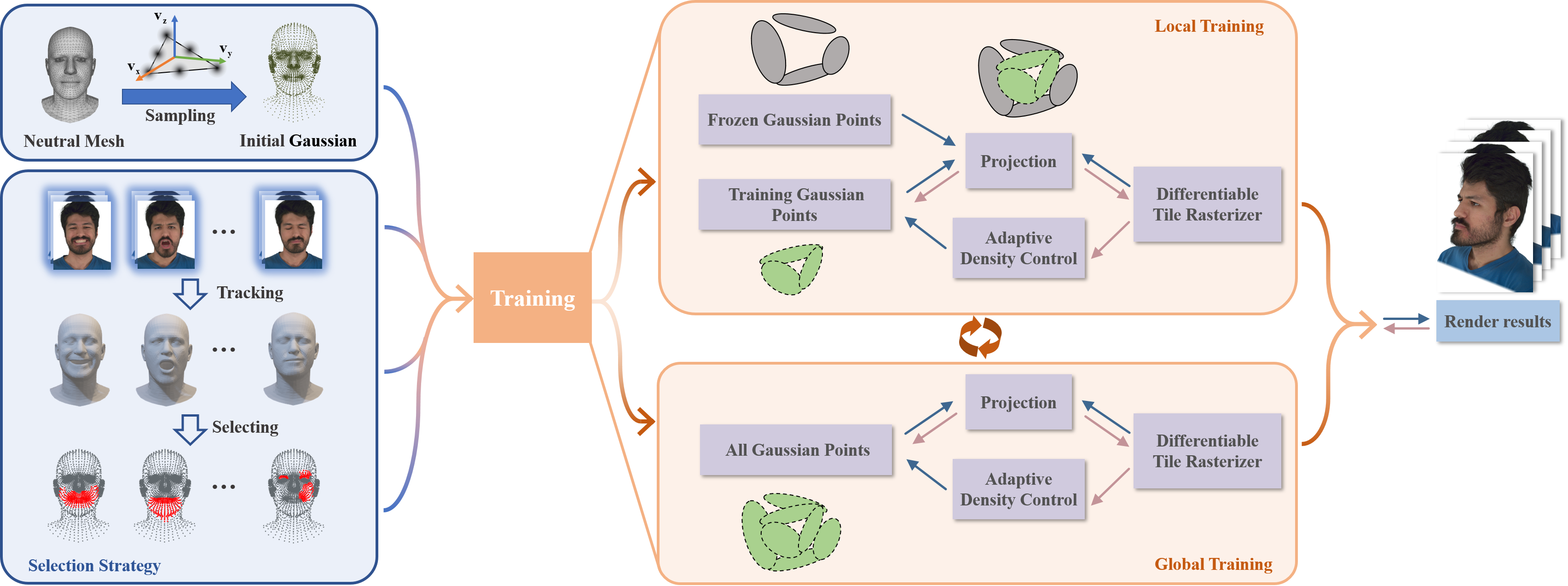}
    \vspace{-0.7cm}
    \caption{Overview of our method. The top-left section illustrates Gaussian initialization, where a local coordinate system is established for each triangle of the neutral mesh of the target head. The bottom-left section shows the FLAME tracking process and the determination of Gaussian splats to be optimized for each frame. The training process consists of two parts: local training and global training. Local training optimizes only the selected Gaussian splats while freezing the others, whereas global training involves all Gaussian splats. These two training modes are alternated to enhance the realism of details while maintaining overall consistency. This process ultimately produces a high-quality dynamic Gaussian model of the head.}
    \label{fig:pipline}
    \vspace{-0.3cm}    
\end{figure*}

Recently, as a point-based 3D scene reconstruction method, 3D Gaussian Splatting has shown better performance than neutral radiation fields in terms of both training speed and rendering speed. Gaussian Splatting, as a new explicit representation method, achieves high efficiency and high fidelity 3D reconstruction. At present, a large number of dynamic human training methods based on 3D Gaussian have been proposed~\cite{psavatar, splattingavatar, animatablegaussians, gaussianavatars, hugs, gart, flashavatar, GauHuman}, including the dynamic Gaussian of human head and the dynamic Gaussian of body. Gaussian avatars~\cite{gaussianavatars} is a FLAME model-based dynamic Gaussian head training method, which embedded Gaussian splats on the FLAME mesh model and uses the mesh model to drive the Gaussian model. PSAvatar~\cite{psavatar} also uses the FLAME mesh model-driven idea, which achieves more accurate driving effects for hair and other positions by diffusing Gaussian splats along the triangular normal direction. ~\cite{relightable} is a method of Gaussian head relighting. ~\cite{gaussianblendshapes} blended expressed Gaussian head through blendshape. HUGS~\cite{hugs} uses SMPL mesh model as the embedded object of Gaussian splats and combines triplane to improve the training effect of the method on monocular video. SplattingAvatar~\cite{splattingavatar} uses the barycentric coordinates and the displacement along the normal of the triangular plane to determine the correspondence between the Gaussian model and the mesh model. Gaussian Head Avatar~\cite{gaussianheadavatar} employs an MLP to fit Gaussian parameters, achieving more realistic results. However, this approach requires extensive training time, with a single Gaussian head taking 1–2 days to train. Currently, most of the training methods have adopted the mesh-based driving method. Each Gaussian splat is embedded in the corresponding triangular face of the mesh model and moves along the corresponding triangular face, so as to realize the animation of the Gaussian model. These methods default to optimizing all 3D Gaussian parameters for each frame during training, which requires a significant amount of GPU memory and considerable training time. Most of the methods need several hours or even days of training time. In addition, such training can result in insufficient optimization of detailed regions, which in turn reduces the realism of the rendered results. Our new method only optimizes some Gaussian splats with higher priority for each frame and freezes the parameters of other Gaussian splats when training dynamic Gaussian, which can effectively improve the quality of details in the rendered results and enhance training efficiency.

\section{Method}

As shown in Figure ~\ref{fig:pipline}, the training dataset consists of multi-view facial expression videos and the corresponding FLAME mesh models for each frame, which includes neutral expression data and other expression data of the target head. The first step is to embed Gaussian splats into the mesh model. We establish a local coordinate system for each triangular face and initialize the Gaussian point cloud on each one. Next, we get the triangular faces with significant deformation across different expressions. For each frame of training, Gaussian points embedded on these faces are selected to participate in the training process. We optimize the position, rotation, scale, opacity, and color of these Gaussian points, while freezing these parameters for the other Gaussian splats. Ultimately, the trained dynamic Gaussian head is obtained.

\subsection{Preliminary}
3D Gaussian is an explicit 3D representation method that utilizes multi-view images of the target object along with camera parameters for training. The target object is represented by many Gaussian splats, where the parameters of each Gaussian splat consist of position $\mu$, covariance matrix $\Sigma$, and color $c$. The Gaussian function is represented as follows:
\begin{equation}
  G(x) = {e^{ - \frac{1}{2}{{(x - \mu )}^T}\mathbf{\Sigma} (x - \mu )}}
  \label{eq:Gx}
\end{equation}
The covariance matrix must be positive semidefinite. Therefore, it is decomposed into two components: a matrix $R$ that represents rotation and a matrix $S$ that represents scaling. To ensure that the covariance matrix remains positive semidefinite during the gradient descent process, it is defined as
\begin{equation}
    \mathbf{\Sigma}  = \mathbf{R}\mathbf{S}{\mathbf{S}^T}{\mathbf{R}^T}
  \label{eq:Sigma}
\end{equation}

For 2D image rendering, the color value $c_i$ of each Gaussian at the pixel location is calculated using spherical harmonic functions, while the weight $\alpha_i$ is computed based on opacity and a 2D Gaussian distribution. The pixel color $C$ is obtained by layering $N$ 2D Gaussians in order of depth:
\begin{equation}
  C = \sum\limits_{i = 1} {{c_i}{\alpha _i}} \prod\limits_{j = 1}^{i - 1} {(1 - {\alpha _j})} 
  \label{eq:C}
\end{equation}

\vspace{-0.5cm}
\subsection{Gaussian Initialization}
\begin{figure}
    \centering
    \includegraphics[width=\linewidth]{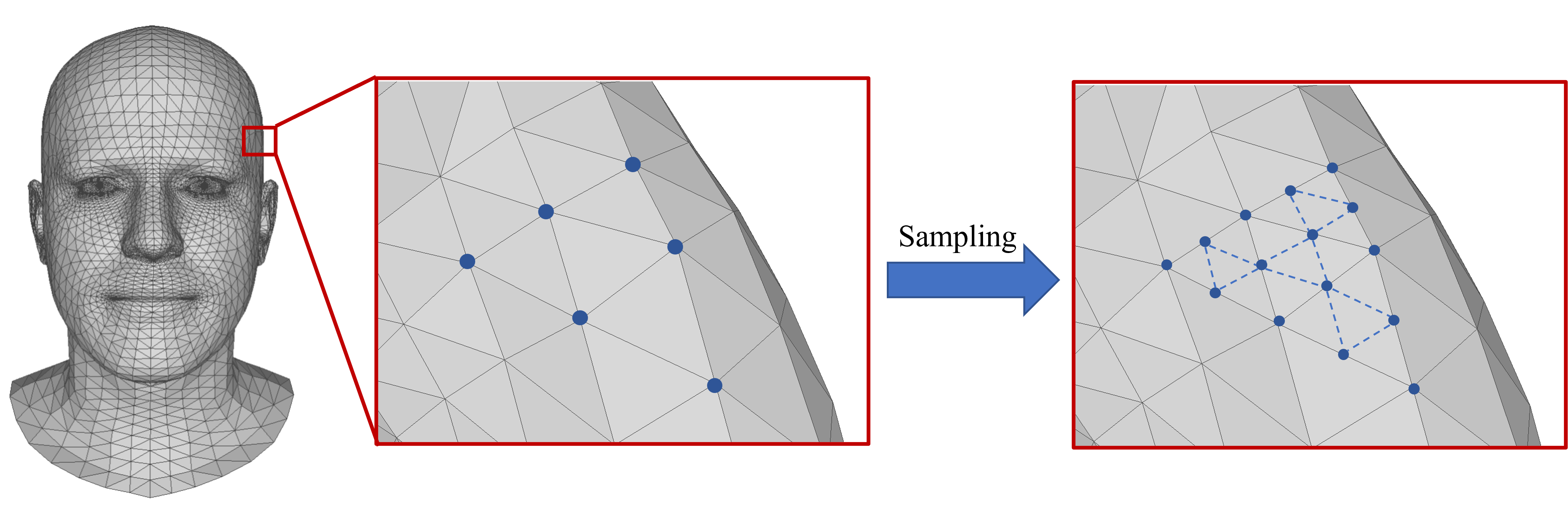}
    \vspace{-0.3cm}
    \caption{Initializing the Gaussian head model method}
    \label{fig:initGS}
\end{figure}

The initialization of the Gaussian point cloud is based on the FLAME mesh model in the neutral expression. The positions of the initial Gaussian point cloud are sampled from the triangular faces of the mesh model. The initial color and opacity are both set to zero. The Gaussian position initialization sampling method is illustrated in Figure~\ref{fig:initGS}. The initial Gaussian splat positions are obtained on the neutral expression FLAME mesh model. For a triangular face of the mesh model, the midpoints of each edge are used to divide it into four smaller triangular faces. The vertices of these four smaller triangular faces, along with the midpoints of each edge, are defined as the world coordinates for the initialization of the Gaussian point cloud. This process ensures that the Gaussian splats are evenly distributed and cover the surface of the triangular face effectively.

\vspace{-0.2cm}
\subsection{Calculating the training splats}

\begin{figure}
    \centering
    \includegraphics[width=\linewidth]{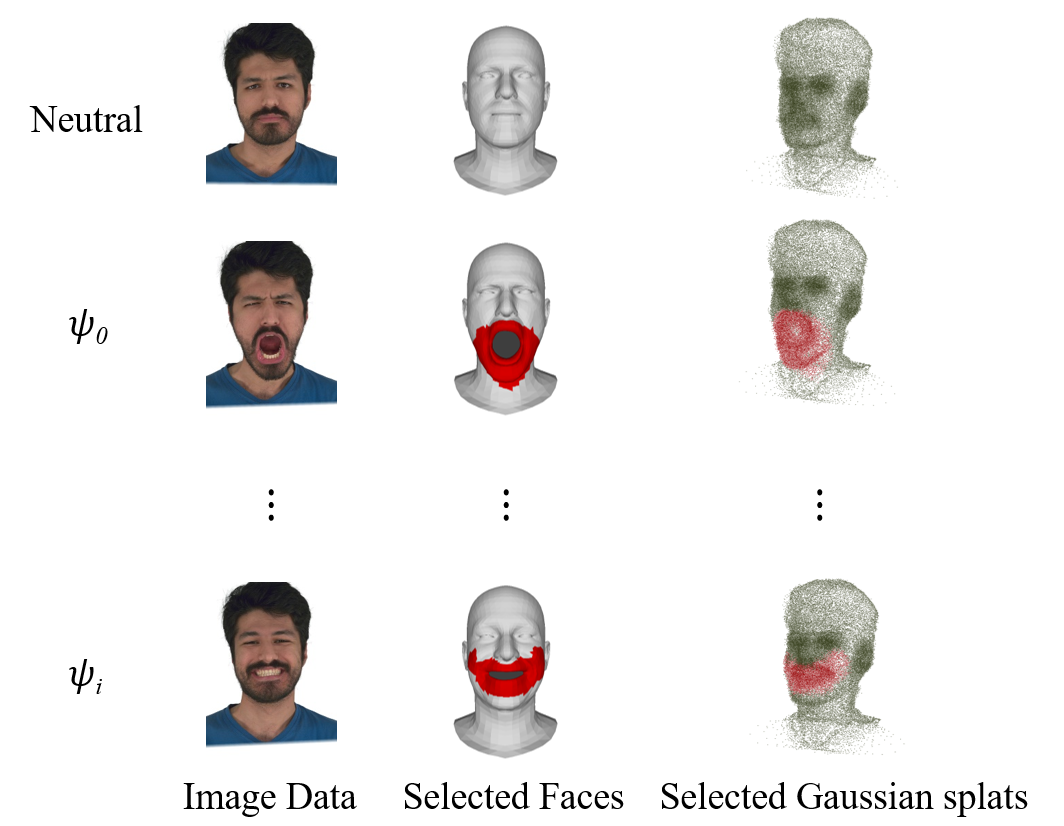}
    \vspace{-0.5cm}
    \caption{Different Gaussian splats are selected for training when training different expressions. $\mathit{\psi}_{0} \dots \mathit{\psi}_{i}$ represent different expressions.}
    \label{fig:selected}
\end{figure}

During the dynamic training of Gaussian splats across different frames, not all Gaussian splats must be involved in the optimization. The Gaussian splats trained in each frame exhibit significant relative changes, varying from frame to frame. It is necessary to determine which Gaussian splats will participate in the optimization for each frame. Since each frame has the FLAME mesh as a prior, we calculate the Gaussian splats that need to be trained for each frame based on the differences in the mesh. The results are shown in Figure~\ref{fig:selected}. This allows us to focus the training on the splats that are most affected by the changes in facial expressions, thereby achieving a more efficient and targeted optimization process.





To prevent global translation and rotation from affecting the computation, the pose parameters of each frame of the FLAME model are set to zero. Using the neutral expression mesh as the reference, the offset distance of each triangle's center in other expressions relative to the neutral mesh's corresponding triangle center is calculated.

\begin{figure}
    \centering
    \includegraphics[width=\linewidth]{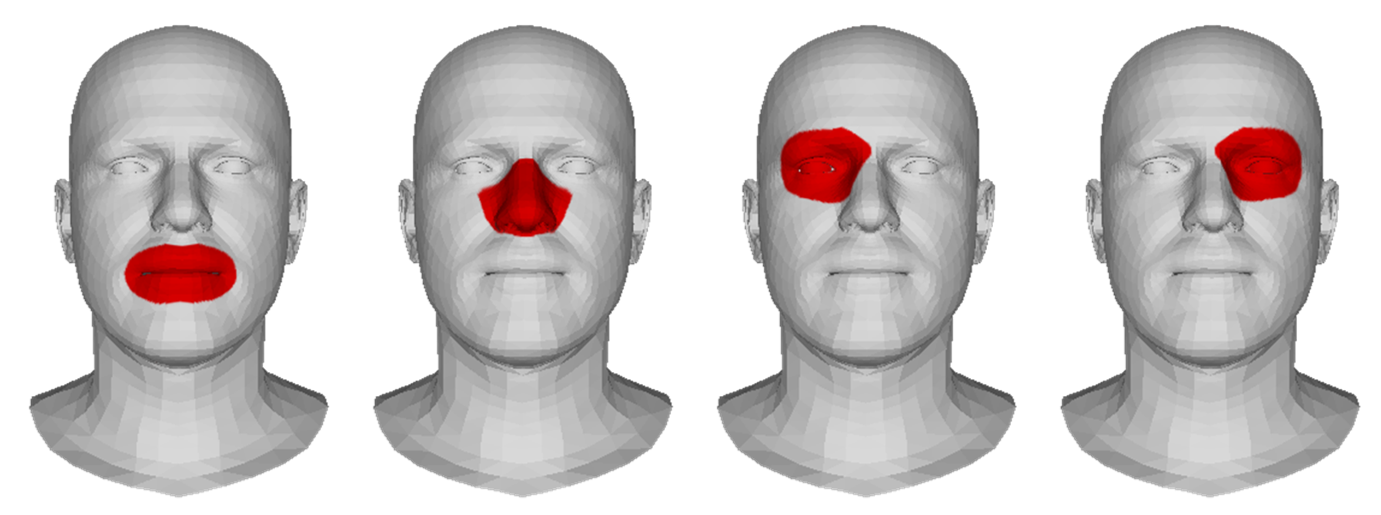}
    \vspace{-0.5cm}
    \caption{Defined key facial regions for Gaussian splat optimization (from left to right: Mouth, Nose, Right Eye, Left Eye}
    \label{fig:region}
\end{figure}

Based on FLAME's Vertex Masks~\cite{flame}, the facial model is divided into several key regions: the left eye, right eye, mouth, and nose. The Figure~\ref{fig:region} illustrates these segmented regions. The Gaussian splats within these regions are optimized as a whole. The key regions are processed first: if the average offset distance of all triangles in a region exceeds a threshold, all the Gaussian splats embedded in those triangles are considered for optimization in that frame. The remaining areas are then processed: if any triangle's offset distance exceeds the threshold, the Gaussian splats in that triangle are also selected for optimization. This approach increases the focus on details during training.

\begin{figure}
    \centering
    \includegraphics[width=\linewidth]{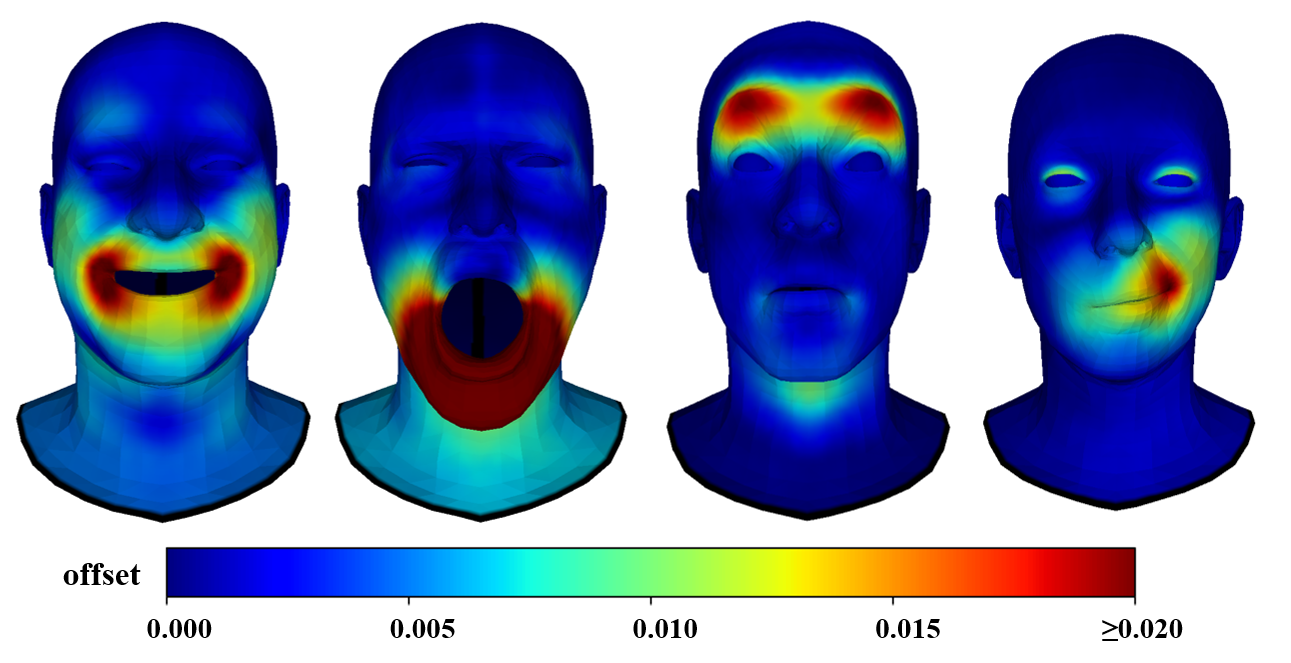}
    \vspace{-0.5cm}
    \caption{Visualization of relative displacements of triangle centers for the FLAME model under different expressions}
    \label{fig:threshold}
\end{figure}

As shown in the Figure~\ref{fig:threshold}, the relative offset results of the triangle centers for the standard-sized FLAME model under different expressions are plotted. It is observed that regions with significant changes have relative offsets greater than 0.01. Therefore, a threshold of 0.01 is chosen for this experiment. 

\vspace{-0.2cm}
\subsection{Gaussian Training Method}
The original Gaussian training method supports only static scenes, making its training results difficult to drive directly. We embed Gaussian splats into the mesh model. This allows the Gaussian splats to move together with the mesh, enabling the driving of the Gaussian model. The key component of the algorithm is establishing the embedding relationship between the 3D mesh and the 3D Gaussian splats. A local coordinate system is established for each triangular face. Each Gaussian splat remains static in the local coordinate space of its parent triangle. As the triangular face translates or rotates in world coordinates, the corresponding Gaussian splats moves accordingly. This enables the driving of Gaussian splats in world space. 

The origin of each local coordinate system is the center position $o$ of the three vertices of the triangle. The vector $v_1$, pointing from the origin to the first vertex, is defined as the x-axis. The normal vector $v_3$ of the triangle serves as the z-axis. The orthogonal vector $v_2$, derived from $v_1$ and $v_3$, becomes the y-axis. This method is applied to establish local coordinate systems for each triangular face, allowing us to obtain their position and rotation information. The rotation matrix for each face is denoted as,
\vspace{-0.1cm}
\begin{equation}
  {\bf{R}} = \left[ {\begin{array}{*{20}{c}}{{v_1}}&{{v_2}}&{{v_3}}\end{array}} \right]
  \vspace{-0.1cm}
  \label{eq:Rv1v2v3}
\end{equation}
    
The local coordinates of each Gaussian splat in the initialized point cloud are calculated. The local positions and the corresponding face indices are then stored in the initialized point cloud.

Additionally, the area of each triangular face is calculated and denoted as $area_t^i$, where $t$ is the current training frame, and $i$ is the triangular face index. The A-pose is defined as the canonical space. The area of each triangular face in the canonical space mesh model is denoted as $area_{can}^i$ The scaling factor $k_t^i$ under different motions is calculated using the following formula:
\vspace{-0.1cm}
\begin{equation}
  k_t^i = \sqrt {area_t^i/area_{can}^i}
  \vspace{-0.1cm}
  \label{eq:k}
\end{equation}

For each 3D Gaussian embedded in a triangular face, we define its local position as $xyz_{em}$, local rotation as $rot_{em}$, and local scaling as $scale_{em}$. Using the parameters obtained above, the Gaussian parameters in world coordinates for each frame can be calculated.
\vspace{-0.1cm}
\begin{equation}
  xy{z_{world}} = k \cdot {\bf{R}} \cdot xy{z_{em}} + o
  \vspace{-0.1cm}
  \label{eq:xyz}
\end{equation}
\begin{equation}
  ro{t_{world}} = {\bf{R}} \cdot ro{t_{em}}
  \vspace{-0.1cm}
  \label{eq:rot}
\end{equation}
\begin{equation}
  scal{e_{world}} = k \cdot scal{e_{em}}
  \vspace{-0.1cm}
  \label{eq:scale}
\end{equation}

The opacity and spherical harmonic parameters are set to remain unchanged across different frames. Based on the results calculated in the previous section, only the selected Gaussian splat parameters are optimized during training for each frame, while all parameters of the other Gaussian splats are frozen. During the Gaussian rasterization step, the training Gaussians and frozen Gaussians are combined to render the target-view image, which is then compared against the ground truth. 

To further enhance the training stability and ensure global consistency across frames, a global optimization step is introduced. After every 20 iterations, all Gaussian splats parameters are temporarily unfrozen and jointly optimized to refine the overall consistency of the Gaussian representation. This step minimizes the global inconsistencies that can arise due to selective optimization.

\subsection{Batch Training}
Dynamic Gaussian training requires a large amount of image data. Typically, a large number of frames with different expressions are necessary to make the Gaussian model driving results more realistic. If all the images are loaded at once, it would consume a significant amount of GPU memory. Reducing image resolution would significantly degrade the quality of training results. 

Therefore, we adopt a batch loading approach during training. In each batch, all viewpoint data for a single frame are loaded, and the position, rotation, and other parameters of the frozen Gaussian splats in the world coordinate system for that frame are computed. During each training iteration, only the parameters of the training Gaussian splats are computed in the world coordinate system. The results are then merged with the frozen Gaussian splats for rendering. Once all the viewpoints of the current frame have been trained, the full set of viewpoint images is loaded for a randomly selected next frame. We use multi-threading to asynchronously load images in a separate thread, which improves the efficiency of the program.

In practice, this approach allows GPU memory usage to be controlled at approximately 3GB, regardless of the dataset size.

\subsection{Optimization and Loss Function}
We refer to the original Gaussian method, and the image loss function is defined as the combination of L1 loss and D-SSIM loss:
\begin{equation}
    {{\mathcal  L}_{{\rm{rgb }}}} = (1 - \lambda ){{\mathcal  L}_1} + \lambda {{\mathcal  L}_{{\rm{D - SSIM }}}}
  \label{eq:loss}
\end{equation}

where $\lambda=0.2$. In addition, we found that additional constraints on Gaussian scaling are needed; otherwise, the results would produce many spiky Gaussian spheres, affecting the driving effect. Therefore, we incorporated a loss function for the Gaussian scale.
\begin{equation}
    {{\mathcal  L}_{{\rm{scaling}}}} = {\left\| {{\rm{ReLU}}(s - \sqrt {2are{a_p}} )} \right\|_2}
    \label{eq:loss_scale}
\end{equation}

In this equation, $s$ refers to the scaling parameter of the Gaussian splats. $area_p$ represents the area of the triangular face associated with each Gaussian splat.

In addition, the premise of Gaussian embedding is that the three-dimensional Gaussian splats should fit the corresponding mesh model. For example, Gaussian splats representing the nose region should not be embedded in the triangular faces of the cheek. Even if the initialized Gaussian splats are positioned near the centers of the correct triangular faces and new Gaussian splats are added nearby, there is no guarantee that the Gaussian splats will remain close to the embedded triangles after optimization. To address this issue, we periodically reset the Gaussian splats that deviate too far from the embedded triangles. During the optimization process, we reset the splat whose distance from the origin of the local coordinate system exceeds $\sqrt {2are{a_p}}$ to ensure the correctness of the embedded Gaussian. 

\section{Experience}
\subsection{Setup}
We utilized the NeRSemble dataset that had been preprocessed by the Gaussianavatars~\cite{gaussianavatars}. The dataset includes the FLAME model for each frame, with vertex offsets added for the hair and teeth regions. Each subject in the dataset comprises a sequence of head frames from 16 different viewpoints, with a resolution of 802×550 pixels. We conducted our training using an NVIDIA RTX 3090 graphics card. During the training process, images from 15 viewpoints were utilized. The frontal viewpoint was excluded from training and instead used for testing to evaluate the results of novel viewpoint generation. We selected two Gaussian head avatar algorithms for comparative analysis: one is the network-driven Gaussian head avatar~\cite{gaussianheadavatar}, and the other is the mesh-driven Gaussianavatars~\cite{gaussianavatars}. 

\subsection{Comparing}
\begin{figure}
    \includegraphics[width=\linewidth]{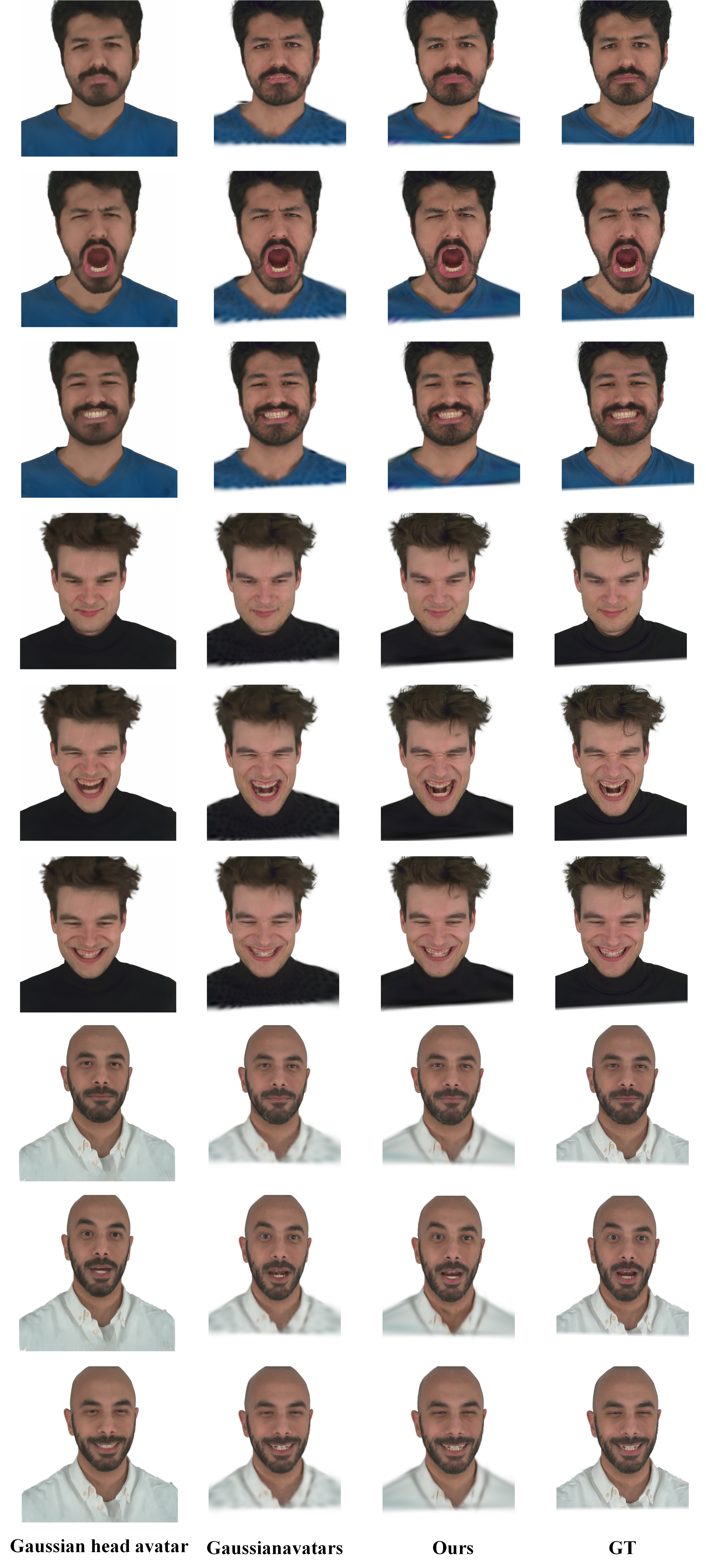}
    \caption{A compare of rendering quality from a new perspective. Our method has achieved superior results in terms of training time and training efficacy. In terms of facial details such as eyes and teeth, our method has outperformed the baseline.}
    \label{fig:compare}
\end{figure}

We tested 3 head datasets with subjects 074, 104, and 218. Since the training process for the Gaussian Head Avatar is divided into two steps, mesh training and Gaussian training, we recorded the training time for both parts. The Gaussian Head Avatar, Gaussian Avatars, and our method were trained for 270 minutes, 10 minutes, and 10 minutes, respectively. The training results of different methods are illustrated in the Figure~\ref{fig:compare}.

The Gaussian Head Avatar method incorporates seven MLP networks, resulting in a computationally intensive optimization process that requires a significant amount of training time to achieve realistic results. Network-based methods, in general, demand considerable time and struggle to produce highly realistic rendered images in a short period. In our experiments, we spent 30 minutes completing the mesh training phase of the Gaussian Head Avatar and an additional 240 minutes on the Gaussian training phase. The rendered images from this short training period still exhibited many unrealistic aspects. In contrast, our method can train a superior Gaussian model for the head in a shorter time.

For mesh-based training methods, both Gaussian Avatars and our approach achieve higher-quality rendered results within a 10-minute training period. Obviously, our method delivers better performance in detailed areas such as teeth and facial wrinkles.

We evaluated the novel viewpoint rendering results using PSNR, SSIM, and LPIPS metrics, with the specific results shown in the Table ~\ref{tab:compare}. It can be observed that our method generally outperforms the other two methods in overall image quality. For subject 218, although the results are comparable to those of Gaussian Avatars, our method achieves more realistic details as shown in the figure.

\begin{table}[h]
    \centering
    \begin{tabular}{clrrr}
        \toprule
        Subject & Method & PSNR$\uparrow$ & SSIM$\uparrow$  & LPIPS$\downarrow$  \\
        \midrule
            & Gaussian head avatar & 27.83     & 0.83      & 0.098  \\
        074 & Gaussain avatars     & 26.07     & 0.87      & 0.13  \\
            & Ours                 & \textbf{28.93}& \textbf{0.88 }& \textbf{0.068}  \\
        \midrule
            & Gaussian head avatar & 23.64     & 0.76      & 0.12  \\
        104 & Gaussain avatars     & 24.81     & 0.85      & 0.14  \\
            & Ours                 & \textbf{27.98}& \textbf{0.90}& \textbf{0.080}  \\
        \midrule
            & Gaussian head avatar & 22.02    & 0.78      & 0.07  \\
        218 & Gaussain avatars     & 28.7     & 0.91      & \textbf{0.069} \\
            & Ours                 & \textbf{29.49 }   & \textbf{0.92}     & 0.074  \\
       
        \bottomrule
    \end{tabular}
    \caption{Nersemble compare table}
    \label{tab:compare}
\end{table}

\subsection{Cross-Identity Reenactment}
\begin{figure}[h]
    \centering
    \includegraphics[width=\linewidth]{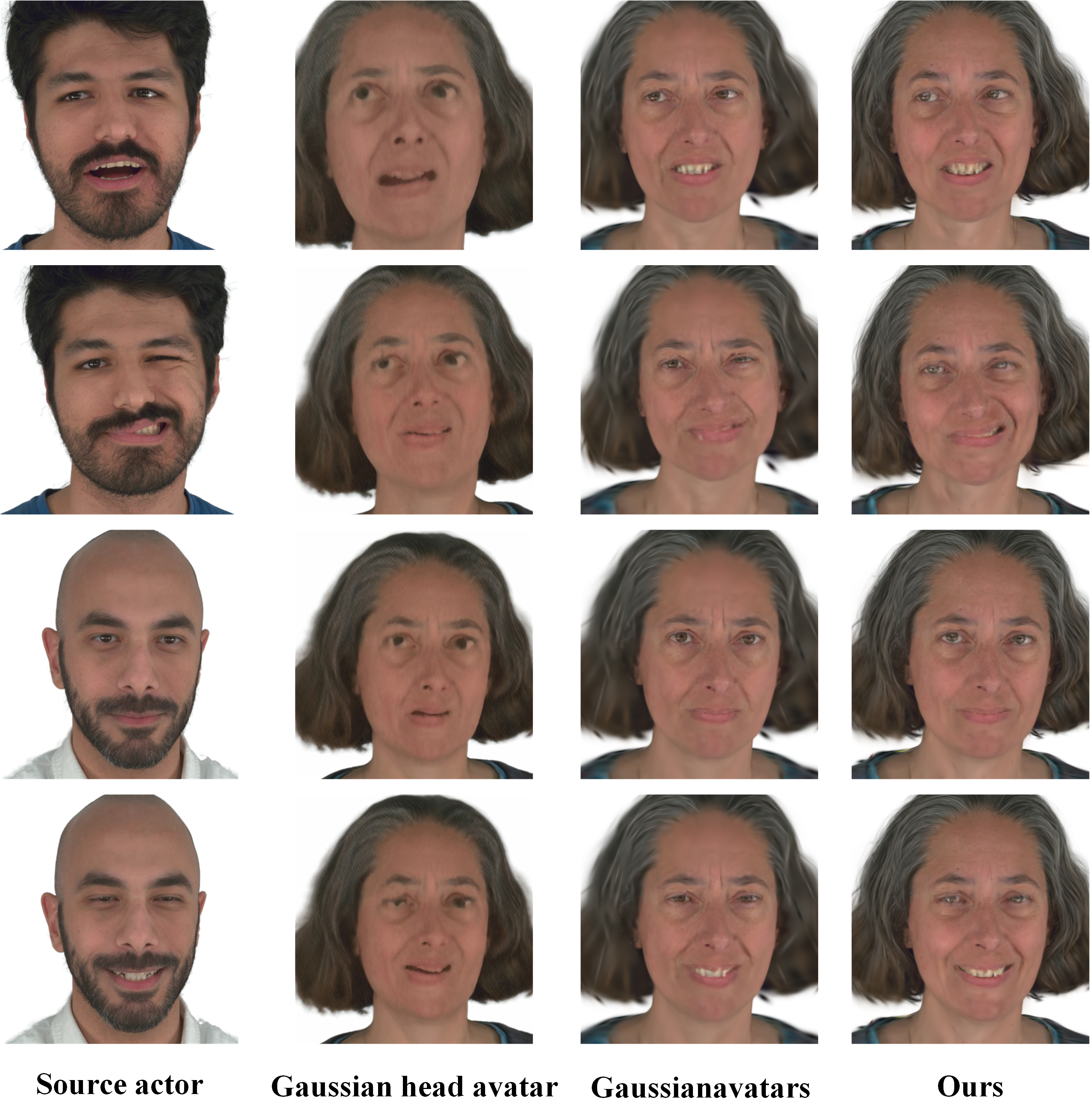}
    \caption{Cross-Identity reenaction images. We used subject 304's own shape parameters and drove its training results using the expression and pose parameters of subjects 074 and 104, respectively. }
    \label{fig:reenaction}
\end{figure}

We performed Cross-identity Reenactment using the expressions of subjects 074 and 218. Following the previously defined training times, the training results of subject 304 were used as the driving target, with some results shown in the Figure~\ref{fig:reenaction}. Due to the high computational cost of training, the Gaussian Head Avatar struggles to achieve realistic results within a short time. Gaussian Avatars produces coarse results in finer areas, such as hair, and generates some artifacts. In comparison, our method delivers more realistic driving results, with details such as wrinkles being more pronounced.

\subsection{Ablation Study}

\begin{figure}[h]
    \centering
    \includegraphics[width=\linewidth]{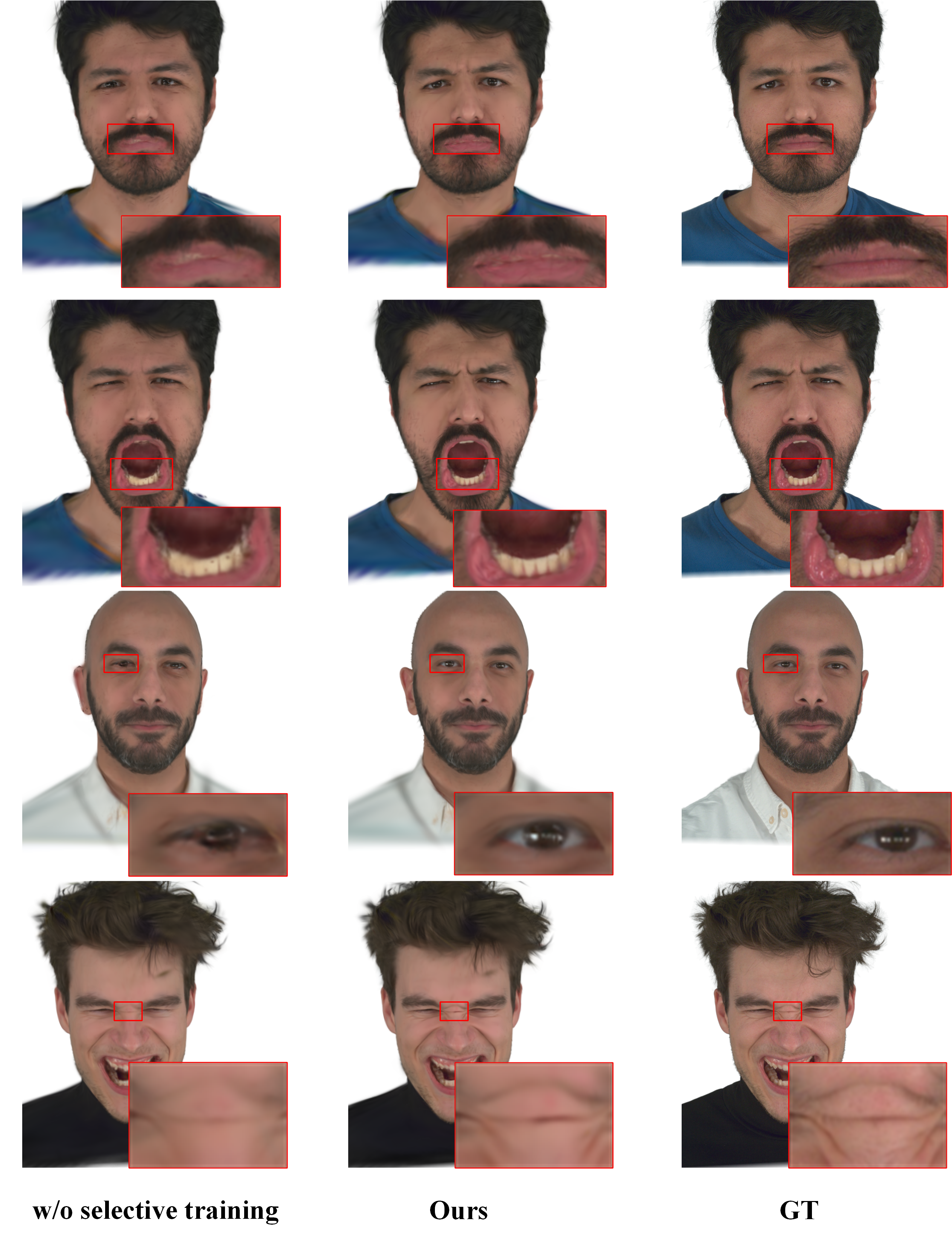}
    \caption{After adding the selective training part, under the same time, the part with larger changes in the head can train better results}
    \label{fig:ablation}
\end{figure}

To validate the effectiveness of the selective optimization training method, we conducted ablation experiments comparing results with and without this approach. The training results under the same time conditions are shown in the Figure~\ref{fig:ablation}.  

Direct global optimization may lead to artifacts such as mesh penetration in certain expressions. Additionally, details in regions like teeth, eyes, and wrinkles tend to be smoothed out, failing to accurately reflect the ground truth. In contrast, our method enhances the representation of fine details, producing Gaussian head models that are closer to the ground truth.  

A comparison of metrics is provided in the table~\ref{tab:Ablation}, demonstrating that incorporating selective training improves the overall quality of the training results.

\begin{table}[h]
    \centering
    \begin{tabular}{lrrr}
        \toprule
        Method & PSNR$\uparrow$ & SSIM$\uparrow$  & LPIPS$\downarrow$  \\
        \midrule
        Ours                    & \textbf{29.49}        & \textbf{0.92}    & \textbf{0.041}     \\
        w/o selective training  & 23.38        & 0.86    & 0.052    \\
       
        \bottomrule
    \end{tabular}
    \caption{Ablation experience table}
    \label{tab:Ablation}
\end{table}

\section{CONCLUSION}

We propose a Gaussian training strategy for selective training. We present a mesh-based method for selecting Gaussian splats to be optimized. Our method effectively enhances the rendering quality of Gaussian head models in detailed regions. We compared our approach with state-of-the-art mesh-based and network-based methods. The results demonstrate that our method can train more realistic Gaussian head models in less time.

\textbf{Limitations: }Since the FLAME model itself does not include information about hair, teeth, and other features, creating accurate training data requires significant effort. The experiments in this paper utilized datasets created by Qian et al.~\cite{gaussianavatars}. However, when testing with other datasets, it is recommended to adjust the vertices for hair, teeth, and similar regions in the FLAME model to achieve optimal training results. Future work could explore using models like NHA~\cite{NHA} or DMTet~\cite{dmtet} as the base mesh for training, enabling automatic fitting of shapes for hair and other features.

\bibliographystyle{ACM-Reference-Format}
\bibliography{sample-bibliography}


\begin{thebibliography}{39}


\ifx \showCODEN    \undefined \def \showCODEN     #1{\unskip}     \fi
\ifx \showDOI      \undefined \def \showDOI       #1{#1}\fi
\ifx \showISBNx    \undefined \def \showISBNx     #1{\unskip}     \fi
\ifx \showISBNxiii \undefined \def \showISBNxiii  #1{\unskip}     \fi
\ifx \showISSN     \undefined \def \showISSN      #1{\unskip}     \fi
\ifx \showLCCN     \undefined \def \showLCCN      #1{\unskip}     \fi
\ifx \shownote     \undefined \def \shownote      #1{#1}          \fi
\ifx \showarticletitle \undefined \def \showarticletitle #1{#1}   \fi
\ifx \showURL      \undefined \def \showURL       {\relax}        \fi
\providecommand\bibfield[2]{#2}
\providecommand\bibinfo[2]{#2}
\providecommand\natexlab[1]{#1}
\providecommand\showeprint[2][]{arXiv:#2}

\bibitem[An et~al\mbox{.}(2023)]%
        {panohead}
\bibfield{author}{\bibinfo{person}{Sizhe An}, \bibinfo{person}{Hongyi Xu}, \bibinfo{person}{Yichun Shi}, \bibinfo{person}{Guoxian Song}, \bibinfo{person}{Umit~Y. Ogras}, {and} \bibinfo{person}{Linjie Luo}.} \bibinfo{year}{2023}\natexlab{}.
\newblock \showarticletitle{PanoHead: Geometry-Aware 3D Full-Head Synthesis in 360deg}. In \bibinfo{booktitle}{\emph{Proceedings of the IEEE/CVF Conference on Computer Vision and Pattern Recognition (CVPR)}}. \bibinfo{pages}{20950--20959}.
\newblock


\bibitem[Dan{\v{e}}{\v{c}}ek et~al\mbox{.}(2022)]%
        {emoca}
\bibfield{author}{\bibinfo{person}{Radek Dan{\v{e}}{\v{c}}ek}, \bibinfo{person}{Michael~J Black}, {and} \bibinfo{person}{Timo Bolkart}.} \bibinfo{year}{2022}\natexlab{}.
\newblock \showarticletitle{Emoca: Emotion driven monocular face capture and animation}. In \bibinfo{booktitle}{\emph{Proceedings of the IEEE/CVF Conference on Computer Vision and Pattern Recognition}}. \bibinfo{pages}{20311--20322}.
\newblock


\bibitem[Dongye et~al\mbox{.}(2024a)]%
        {GaussianReplacement}
\bibfield{author}{\bibinfo{person}{Xiaonuo Dongye}, \bibinfo{person}{Hanzhi Guo}, \bibinfo{person}{Yihua Bao}, {and} \bibinfo{person}{Dongdong Weng}.} \bibinfo{year}{2024}\natexlab{a}.
\newblock \showarticletitle{Gaussian Replacement: Gaussians-Mesh Joint Rendering for Real-Time VR Interaction}. In \bibinfo{booktitle}{\emph{Chinese Conference on Image and Graphics Technologies}}. Springer, \bibinfo{pages}{312--326}.
\newblock


\bibitem[Dongye et~al\mbox{.}(2024b)]%
        {adaptive}
\bibfield{author}{\bibinfo{person}{Xiaonuo Dongye}, \bibinfo{person}{Hanzhi Guo}, \bibinfo{person}{Haiyan Jiang}, {and} \bibinfo{person}{Dongdong Weng}.} \bibinfo{year}{2024}\natexlab{b}.
\newblock \showarticletitle{Adaptive Levels of Detail for Human Gaussian Splats with Hierarchical Embedding}. In \bibinfo{booktitle}{\emph{2024 IEEE International Symposium on Mixed and Augmented Reality Adjunct (ISMAR-Adjunct)}}. IEEE, \bibinfo{pages}{361--362}.
\newblock


\bibitem[Feng et~al\mbox{.}(2021)]%
        {deca}
\bibfield{author}{\bibinfo{person}{Yao Feng}, \bibinfo{person}{Haiwen Feng}, \bibinfo{person}{Michael~J Black}, {and} \bibinfo{person}{Timo Bolkart}.} \bibinfo{year}{2021}\natexlab{}.
\newblock \showarticletitle{Learning an animatable detailed 3D face model from in-the-wild images}.
\newblock \bibinfo{journal}{\emph{ACM Transactions on Graphics (ToG)}} \bibinfo{volume}{40}, \bibinfo{number}{4} (\bibinfo{year}{2021}), \bibinfo{pages}{1--13}.
\newblock


\bibitem[Grassal et~al\mbox{.}(2022)]%
        {NHA}
\bibfield{author}{\bibinfo{person}{Philip-William Grassal}, \bibinfo{person}{Malte Prinzler}, \bibinfo{person}{Titus Leistner}, \bibinfo{person}{Carsten Rother}, \bibinfo{person}{Matthias Nie{\ss}ner}, {and} \bibinfo{person}{Justus Thies}.} \bibinfo{year}{2022}\natexlab{}.
\newblock \showarticletitle{Neural head avatars from monocular RGB videos}. In \bibinfo{booktitle}{\emph{Proceedings of the IEEE/CVF Conference on Computer Vision and Pattern Recognition}}. \bibinfo{pages}{18653--18664}.
\newblock


\bibitem[Hu and Liu(2023)]%
        {GauHuman}
\bibfield{author}{\bibinfo{person}{Shoukang Hu} {and} \bibinfo{person}{Ziwei Liu}.} \bibinfo{year}{2023}\natexlab{}.
\newblock \showarticletitle{GauHuman: Articulated Gaussian Splatting for Real-Time 3D Human Rendering}.
\newblock \bibinfo{journal}{\emph{arXiv preprint}} (\bibinfo{year}{2023}).
\newblock


\bibitem[Kerbl et~al\mbox{.}(2023)]%
        {gaussian}
\bibfield{author}{\bibinfo{person}{Bernhard Kerbl}, \bibinfo{person}{Georgios Kopanas}, \bibinfo{person}{Thomas Leimk{\"u}hler}, {and} \bibinfo{person}{George Drettakis}.} \bibinfo{year}{2023}\natexlab{}.
\newblock \showarticletitle{3D Gaussian Splatting for Real-Time Radiance Field Rendering.}
\newblock \bibinfo{journal}{\emph{ACM Transactions on Graphics(TOG)}} \bibinfo{volume}{42}, \bibinfo{number}{4} (\bibinfo{year}{2023}), \bibinfo{pages}{1--14}.
\newblock


\bibitem[Kocabas et~al\mbox{.}(2023)]%
        {hugs}
\bibfield{author}{\bibinfo{person}{Muhammed Kocabas}, \bibinfo{person}{Rick Chang}, \bibinfo{person}{James Gabriel}, \bibinfo{person}{Oncel Tuzel}, {and} \bibinfo{person}{Anurag Ranjan}.} \bibinfo{year}{2023}\natexlab{}.
\newblock \bibinfo{title}{HUGS: Human Gaussian Splats}.
\newblock
\newblock
\urldef\tempurl%
\url{https://arxiv.org/abs/2311.17910}
\showURL{%
\tempurl}


\bibitem[Lei et~al\mbox{.}(2024)]%
        {gart}
\bibfield{author}{\bibinfo{person}{Jiahui Lei}, \bibinfo{person}{Yufu Wang}, \bibinfo{person}{Georgios Pavlakos}, \bibinfo{person}{Lingjie Liu}, {and} \bibinfo{person}{Kostas Daniilidis}.} \bibinfo{year}{2024}\natexlab{}.
\newblock \showarticletitle{Gart: Gaussian articulated template models}. In \bibinfo{booktitle}{\emph{Proceedings of the IEEE/CVF Conference on Computer Vision and Pattern Recognition}}. \bibinfo{pages}{19876--19887}.
\newblock


\bibitem[Li et~al\mbox{.}(2017)]%
        {flame}
\bibfield{author}{\bibinfo{person}{Tianye Li}, \bibinfo{person}{Timo Bolkart}, \bibinfo{person}{Michael~J Black}, \bibinfo{person}{Hao Li}, {and} \bibinfo{person}{Javier Romero}.} \bibinfo{year}{2017}\natexlab{}.
\newblock \showarticletitle{Learning a model of facial shape and expression from 4D scans.}
\newblock \bibinfo{journal}{\emph{ACM Trans. Graph.}} \bibinfo{volume}{36}, \bibinfo{number}{6} (\bibinfo{year}{2017}), \bibinfo{pages}{194--1}.
\newblock


\bibitem[Li et~al\mbox{.}(2024)]%
        {animatablegaussians}
\bibfield{author}{\bibinfo{person}{Zhe Li}, \bibinfo{person}{Zerong Zheng}, \bibinfo{person}{Lizhen Wang}, {and} \bibinfo{person}{Yebin Liu}.} \bibinfo{year}{2024}\natexlab{}.
\newblock \showarticletitle{Animatable Gaussians: Learning Pose-dependent Gaussian Maps for High-fidelity Human Avatar Modeling}. In \bibinfo{booktitle}{\emph{Proceedings of the IEEE/CVF Conference on Computer Vision and Pattern Recognition (CVPR)}}.
\newblock


\bibitem[Liu et~al\mbox{.}(2021)]%
        {neuralactor}
\bibfield{author}{\bibinfo{person}{Lingjie Liu}, \bibinfo{person}{Marc Habermann}, \bibinfo{person}{Viktor Rudnev}, \bibinfo{person}{Kripasindhu Sarkar}, \bibinfo{person}{Jiatao Gu}, {and} \bibinfo{person}{Christian Theobalt}.} \bibinfo{year}{2021}\natexlab{}.
\newblock \showarticletitle{Neural Actor: Neural Free-view Synthesis of Human Actors with Pose Control}.
\newblock \bibinfo{journal}{\emph{ACM SIGGRAPH Asia}} (\bibinfo{year}{2021}).
\newblock


\bibitem[Loper et~al\mbox{.}(2023)]%
        {smpl}
\bibfield{author}{\bibinfo{person}{Matthew Loper}, \bibinfo{person}{Naureen Mahmood}, \bibinfo{person}{Javier Romero}, \bibinfo{person}{Gerard Pons-Moll}, {and} \bibinfo{person}{Michael~J Black}.} \bibinfo{year}{2023}\natexlab{}.
\newblock \showarticletitle{SMPL: A skinned multi-person linear model}.
\newblock In \bibinfo{booktitle}{\emph{Seminal Graphics Papers: Pushing the Boundaries, Volume 2}}. \bibinfo{pages}{851--866}.
\newblock


\bibitem[Luo et~al\mbox{.}(2024)]%
        {splatface}
\bibfield{author}{\bibinfo{person}{Jiahao Luo}, \bibinfo{person}{Jing Liu}, {and} \bibinfo{person}{James Davis}.} \bibinfo{year}{2024}\natexlab{}.
\newblock \showarticletitle{SplatFace: Gaussian Splat Face Reconstruction Leveraging an Optimizable Surface}.
\newblock \bibinfo{journal}{\emph{arXiv preprint arXiv:2403.18784}} (\bibinfo{year}{2024}).
\newblock


\bibitem[Ma et~al\mbox{.}(2024)]%
        {gaussianblendshapes}
\bibfield{author}{\bibinfo{person}{Shengjie Ma}, \bibinfo{person}{Yanlin Weng}, \bibinfo{person}{Tianjia Shao}, {and} \bibinfo{person}{Kun Zhou}.} \bibinfo{year}{2024}\natexlab{}.
\newblock \showarticletitle{3D Gaussian Blendshapes for Head Avatar Animation}. In \bibinfo{booktitle}{\emph{ACM SIGGRAPH Conference Proceedings, Denver, CO, United States, July 28 - August 1, 2024}}.
\newblock


\bibitem[Mildenhall et~al\mbox{.}(2021)]%
        {nerf}
\bibfield{author}{\bibinfo{person}{Ben Mildenhall}, \bibinfo{person}{Pratul~P Srinivasan}, \bibinfo{person}{Matthew Tancik}, \bibinfo{person}{Jonathan~T Barron}, \bibinfo{person}{Ravi Ramamoorthi}, {and} \bibinfo{person}{Ren Ng}.} \bibinfo{year}{2021}\natexlab{}.
\newblock \showarticletitle{Nerf: Representing scenes as neural radiance fields for view synthesis}.
\newblock \bibinfo{journal}{\emph{Commun. ACM}} \bibinfo{volume}{65}, \bibinfo{number}{1} (\bibinfo{year}{2021}), \bibinfo{pages}{99--106}.
\newblock


\bibitem[M\"uller et~al\mbox{.}(2022)]%
        {instant}
\bibfield{author}{\bibinfo{person}{Thomas M\"uller}, \bibinfo{person}{Alex Evans}, \bibinfo{person}{Christoph Schied}, {and} \bibinfo{person}{Alexander Keller}.} \bibinfo{year}{2022}\natexlab{}.
\newblock \showarticletitle{Instant Neural Graphics Primitives with a Multiresolution Hash Encoding}.
\newblock \bibinfo{journal}{\emph{ACM Trans. Graph.}} \bibinfo{volume}{41}, \bibinfo{number}{4}, Article \bibinfo{articleno}{102} (\bibinfo{date}{July} \bibinfo{year}{2022}), \bibinfo{numpages}{15}~pages.
\newblock
\urldef\tempurl%
\url{https://doi.org/10.1145/3528223.3530127}
\showDOI{\tempurl}


\bibitem[Pan et~al\mbox{.}(2024)]%
        {humansplat}
\bibfield{author}{\bibinfo{person}{Panwang Pan}, \bibinfo{person}{Zhuo Su}, \bibinfo{person}{Chenguo Lin}, \bibinfo{person}{Zhen Fan}, \bibinfo{person}{Yongjie Zhang}, \bibinfo{person}{Zeming Li}, \bibinfo{person}{Tingting Shen}, \bibinfo{person}{Yadong Mu}, {and} \bibinfo{person}{Yebin Liu}.} \bibinfo{year}{2024}\natexlab{}.
\newblock \showarticletitle{HumanSplat: Generalizable Single-Image Human Gaussian Splatting with Structure Priors}. In \bibinfo{booktitle}{\emph{Advances in Neural Information Processing Systems (NeurIPS)}}.
\newblock


\bibitem[Park et~al\mbox{.}(2020)]%
        {nerfies}
\bibfield{author}{\bibinfo{person}{Keunhong Park}, \bibinfo{person}{Utkarsh Sinha}, \bibinfo{person}{Jonathan~T. Barron}, \bibinfo{person}{Sofien Bouaziz}, \bibinfo{person}{Dan Goldman}, \bibinfo{person}{Steven Seitz}, {and} \bibinfo{person}{Ricardo Martin-Brualla}.} \bibinfo{year}{2020}\natexlab{}.
\newblock \showarticletitle{Deformable Neural Radiance Fields}.
\newblock \bibinfo{journal}{\emph{https://arxiv.org/abs/2011.12948}} (\bibinfo{year}{2020}).
\newblock


\bibitem[Pavlakos et~al\mbox{.}(2019)]%
        {smplx}
\bibfield{author}{\bibinfo{person}{Georgios Pavlakos}, \bibinfo{person}{Vasileios Choutas}, \bibinfo{person}{Nima Ghorbani}, \bibinfo{person}{Timo Bolkart}, \bibinfo{person}{Ahmed~AA Osman}, \bibinfo{person}{Dimitrios Tzionas}, {and} \bibinfo{person}{Michael~J Black}.} \bibinfo{year}{2019}\natexlab{}.
\newblock \showarticletitle{Expressive body capture: 3d hands, face, and body from a single image}. In \bibinfo{booktitle}{\emph{Proceedings of the IEEE/CVF conference on computer vision and pattern recognition}}. \bibinfo{pages}{10975--10985}.
\newblock


\bibitem[Peng et~al\mbox{.}(2021)]%
        {animatablenerf}
\bibfield{author}{\bibinfo{person}{Sida Peng}, \bibinfo{person}{Junting Dong}, \bibinfo{person}{Qianqian Wang}, \bibinfo{person}{Shangzhan Zhang}, \bibinfo{person}{Qing Shuai}, \bibinfo{person}{Xiaowei Zhou}, {and} \bibinfo{person}{Hujun Bao}.} \bibinfo{year}{2021}\natexlab{}.
\newblock \showarticletitle{Animatable neural radiance fields for modeling dynamic human bodies}. In \bibinfo{booktitle}{\emph{Proceedings of the IEEE/CVF International Conference on Computer Vision}}. \bibinfo{pages}{14314--14323}.
\newblock


\bibitem[Pumarola et~al\mbox{.}(2020)]%
        {D_NeRF}
\bibfield{author}{\bibinfo{person}{Albert Pumarola}, \bibinfo{person}{Enric Corona}, \bibinfo{person}{Gerard Pons-Moll}, {and} \bibinfo{person}{Francesc Moreno-Noguer}.} \bibinfo{year}{2020}\natexlab{}.
\newblock \showarticletitle{{D-NeRF}: Neural Radiance Fields for Dynamic Scenes}.
\newblock \bibinfo{journal}{\emph{https://arxiv.org/abs/2011.13961}} (\bibinfo{year}{2020}).
\newblock


\bibitem[Qian et~al\mbox{.}(2024)]%
        {gaussianavatars}
\bibfield{author}{\bibinfo{person}{Shenhan Qian}, \bibinfo{person}{Tobias Kirschstein}, \bibinfo{person}{Liam Schoneveld}, \bibinfo{person}{Davide Davoli}, \bibinfo{person}{Simon Giebenhain}, {and} \bibinfo{person}{Matthias Nie{\ss}ner}.} \bibinfo{year}{2024}\natexlab{}.
\newblock \showarticletitle{Gaussianavatars: Photorealistic head avatars with rigged 3d gaussians}. In \bibinfo{booktitle}{\emph{Proceedings of the IEEE/CVF Conference on Computer Vision and Pattern Recognition}}. \bibinfo{pages}{20299--20309}.
\newblock


\bibitem[Reiser et~al\mbox{.}(2021)]%
        {kilonerf}
\bibfield{author}{\bibinfo{person}{Christian Reiser}, \bibinfo{person}{Songyou Peng}, \bibinfo{person}{Yiyi Liao}, {and} \bibinfo{person}{Andreas Geiger}.} \bibinfo{year}{2021}\natexlab{}.
\newblock \showarticletitle{Kilonerf: Speeding up neural radiance fields with thousands of tiny mlps}. In \bibinfo{booktitle}{\emph{Proceedings of the IEEE/CVF international conference on computer vision}}. \bibinfo{pages}{14335--14345}.
\newblock


\bibitem[Romero et~al\mbox{.}(2022)]%
        {mano}
\bibfield{author}{\bibinfo{person}{Javier Romero}, \bibinfo{person}{Dimitrios Tzionas}, {and} \bibinfo{person}{Michael~J Black}.} \bibinfo{year}{2022}\natexlab{}.
\newblock \showarticletitle{Embodied hands: Modeling and capturing hands and bodies together}.
\newblock \bibinfo{journal}{\emph{arXiv preprint arXiv:2201.02610}} (\bibinfo{year}{2022}).
\newblock


\bibitem[Saito et~al\mbox{.}(2024)]%
        {relightable}
\bibfield{author}{\bibinfo{person}{Shunsuke Saito}, \bibinfo{person}{Gabriel Schwartz}, \bibinfo{person}{Tomas Simon}, \bibinfo{person}{Junxuan Li}, {and} \bibinfo{person}{Giljoo Nam}.} \bibinfo{year}{2024}\natexlab{}.
\newblock \showarticletitle{Relightable gaussian codec avatars}. In \bibinfo{booktitle}{\emph{Proceedings of the IEEE/CVF Conference on Computer Vision and Pattern Recognition}}. \bibinfo{pages}{130--141}.
\newblock


\bibitem[Shao et~al\mbox{.}(2024)]%
        {splattingavatar}
\bibfield{author}{\bibinfo{person}{Zhijing Shao}, \bibinfo{person}{Zhaolong Wang}, \bibinfo{person}{Zhuang Li}, \bibinfo{person}{Duotun Wang}, \bibinfo{person}{Xiangru Lin}, \bibinfo{person}{Yu Zhang}, \bibinfo{person}{Mingming Fan}, {and} \bibinfo{person}{Zeyu Wang}.} \bibinfo{year}{2024}\natexlab{}.
\newblock \showarticletitle{{SplattingAvatar: Realistic Real-Time Human Avatars with Mesh-Embedded Gaussian Splatting}}. In \bibinfo{booktitle}{\emph{Proceedings of the IEEE/CVF Conference on Computer Vision and Pattern Recognition (CVPR)}}.
\newblock


\bibitem[Shen et~al\mbox{.}(2021)]%
        {dmtet}
\bibfield{author}{\bibinfo{person}{Tianchang Shen}, \bibinfo{person}{Jun Gao}, \bibinfo{person}{Kangxue Yin}, \bibinfo{person}{Ming-Yu Liu}, {and} \bibinfo{person}{Sanja Fidler}.} \bibinfo{year}{2021}\natexlab{}.
\newblock \showarticletitle{Deep Marching Tetrahedra: a Hybrid Representation for High-Resolution 3D Shape Synthesis}. In \bibinfo{booktitle}{\emph{Advances in Neural Information Processing Systems (NeurIPS)}}.
\newblock


\bibitem[Wen et~al\mbox{.}(2024)]%
        {gomavatar}
\bibfield{author}{\bibinfo{person}{Jing Wen}, \bibinfo{person}{Xiaoming Zhao}, \bibinfo{person}{Zhongzheng Ren}, \bibinfo{person}{Alex Schwing}, {and} \bibinfo{person}{Shenlong Wang}.} \bibinfo{year}{2024}\natexlab{}.
\newblock \showarticletitle{{GoMAvatar: Efficient Animatable Human Modeling from Monocular Video Using Gaussians-on-Mesh}}. In \bibinfo{booktitle}{\emph{CVPR}}.
\newblock


\bibitem[Weng et~al\mbox{.}(2022)]%
        {humannerf}
\bibfield{author}{\bibinfo{person}{Chung-Yi Weng}, \bibinfo{person}{Brian Curless}, \bibinfo{person}{Pratul~P. Srinivasan}, \bibinfo{person}{Jonathan~T. Barron}, {and} \bibinfo{person}{Ira Kemelmacher-Shlizerman}.} \bibinfo{year}{2022}\natexlab{}.
\newblock \showarticletitle{Human{N}e{RF}: Free-Viewpoint Rendering of Moving People From Monocular Video}. In \bibinfo{booktitle}{\emph{Proceedings of the IEEE/CVF Conference on Computer Vision and Pattern Recognition (CVPR)}}. \bibinfo{pages}{16210--16220}.
\newblock


\bibitem[Xiang et~al\mbox{.}(2024)]%
        {flashavatar}
\bibfield{author}{\bibinfo{person}{Jun Xiang}, \bibinfo{person}{Xuan Gao}, \bibinfo{person}{Yudong Guo}, {and} \bibinfo{person}{Juyong Zhang}.} \bibinfo{year}{2024}\natexlab{}.
\newblock \showarticletitle{FlashAvatar: High-fidelity Head Avatar with Efficient Gaussian Embedding}. In \bibinfo{booktitle}{\emph{The IEEE Conference on Computer Vision and Pattern Recognition (CVPR)}}.
\newblock


\bibitem[Xiao et~al\mbox{.}(2024)]%
        {bridging}
\bibfield{author}{\bibinfo{person}{Yuting Xiao}, \bibinfo{person}{Xuan Wang}, \bibinfo{person}{Jiafei Li}, \bibinfo{person}{Hongrui Cai}, \bibinfo{person}{Yanbo Fan}, \bibinfo{person}{Nan Xue}, \bibinfo{person}{Minghui Yang}, \bibinfo{person}{Yujun Shen}, {and} \bibinfo{person}{Shenghua Gao}.} \bibinfo{year}{2024}\natexlab{}.
\newblock \showarticletitle{Bridging 3D Gaussian and Mesh for Freeview Video Rendering}.
\newblock \bibinfo{journal}{\emph{arXiv preprint arXiv:2403.11453}} (\bibinfo{year}{2024}).
\newblock


\bibitem[Xie et~al\mbox{.}(2023)]%
        {physgaussian}
\bibfield{author}{\bibinfo{person}{Tianyi Xie}, \bibinfo{person}{Zeshun Zong}, \bibinfo{person}{Yuxing Qiu}, \bibinfo{person}{Xuan Li}, \bibinfo{person}{Yutao Feng}, \bibinfo{person}{Yin Yang}, {and} \bibinfo{person}{Chenfanfu Jiang}.} \bibinfo{year}{2023}\natexlab{}.
\newblock \showarticletitle{PhysGaussian: Physics-Integrated 3D Gaussians for Generative Dynamics}.
\newblock \bibinfo{journal}{\emph{arXiv preprint arXiv:2311.12198}} (\bibinfo{year}{2023}).
\newblock


\bibitem[Xu et~al\mbox{.}(2024)]%
        {gaussianheadavatar}
\bibfield{author}{\bibinfo{person}{Yuelang Xu}, \bibinfo{person}{Benwang Chen}, \bibinfo{person}{Zhe Li}, \bibinfo{person}{Hongwen Zhang}, \bibinfo{person}{Lizhen Wang}, \bibinfo{person}{Zerong Zheng}, {and} \bibinfo{person}{Yebin Liu}.} \bibinfo{year}{2024}\natexlab{}.
\newblock \showarticletitle{Gaussian Head Avatar: Ultra High-fidelity Head Avatar via Dynamic Gaussians}. In \bibinfo{booktitle}{\emph{Proceedings of the IEEE/CVF Conference on Computer Vision and Pattern Recognition (CVPR)}}.
\newblock


\bibitem[Ye et~al\mbox{.}(2024)]%
        {gaussian_grouping}
\bibfield{author}{\bibinfo{person}{Mingqiao Ye}, \bibinfo{person}{Martin Danelljan}, \bibinfo{person}{Fisher Yu}, {and} \bibinfo{person}{Lei Ke}.} \bibinfo{year}{2024}\natexlab{}.
\newblock \showarticletitle{Gaussian Grouping: Segment and Edit Anything in 3D Scenes}. In \bibinfo{booktitle}{\emph{ECCV}}.
\newblock


\bibitem[Yu et~al\mbox{.}(2021)]%
        {plenoctrees}
\bibfield{author}{\bibinfo{person}{Alex Yu}, \bibinfo{person}{Ruilong Li}, \bibinfo{person}{Matthew Tancik}, \bibinfo{person}{Hao Li}, \bibinfo{person}{Ren Ng}, {and} \bibinfo{person}{Angjoo Kanazawa}.} \bibinfo{year}{2021}\natexlab{}.
\newblock \showarticletitle{Plenoctrees for real-time rendering of neural radiance fields}. In \bibinfo{booktitle}{\emph{Proceedings of the IEEE/CVF International Conference on Computer Vision}}. \bibinfo{pages}{5752--5761}.
\newblock


\bibitem[Zhao et~al\mbox{.}(2024)]%
        {psavatar}
\bibfield{author}{\bibinfo{person}{Zhongyuan Zhao}, \bibinfo{person}{Zhenyu Bao}, \bibinfo{person}{Qing Li}, \bibinfo{person}{Guoping Qiu}, {and} \bibinfo{person}{Kanglin Liu}.} \bibinfo{year}{2024}\natexlab{}.
\newblock \showarticletitle{Psavatar: A point-based morphable shape model for real-time head avatar creation with 3d gaussian splatting}.
\newblock \bibinfo{journal}{\emph{arXiv preprint arXiv:2401.12900}} (\bibinfo{year}{2024}).
\newblock


\bibitem[Zielonka et~al\mbox{.}(2022)]%
        {mica}
\bibfield{author}{\bibinfo{person}{Wojciech Zielonka}, \bibinfo{person}{Timo Bolkart}, {and} \bibinfo{person}{Justus Thies}.} \bibinfo{year}{2022}\natexlab{}.
\newblock \showarticletitle{Towards metrical reconstruction of human faces}. In \bibinfo{booktitle}{\emph{European conference on computer vision}}. Springer, \bibinfo{pages}{250--269}.
\newblock


\end{thebibliography}

\end{document}